\documentstyle[prl,aps,epsf]{revtex}

\begin{document}
\def \beq {\begin{equation}}
\def \eeq {\end{equation}}
\def \bes {\begin{eqnarray}}
\def \ees {\end{eqnarray}}
\def \ni {\noindent}
\def \nn {\nonumber}
\def \z {\tilde{z}}

\title{
Demonstration of the Lateral Casimir Force}

\author{
F.~Chen,
U.~Mohideen\footnote{Corresponding author.
E-mail: umar.mohideen@ucr.edu},
}
\address
{Department of Physics, University of California,
Riverside, California 92521\\
}

\author{
G.~L.~Klimchitskaya\footnote{On leave from
North-West Polytechnical University,
\\St.Petersburg, Russia.}
and V.~M.~Mostepanenko\footnote{On leave from
Research and Innovation Enterprise ``Modus'',
\\Moscow, Russia.}
}
\address
{Departamento de F\'{\i}sica, Universidade Federal da Para\'{\i}ba,
C.P.~5008, CEP 58059-970,
Jo\~{a}o Pessoa, Pb-Brazil
}
\maketitle

{\abstract{The lateral Casimir force between a sinusoidally
corrugated gold coated plate and large sphere was measured
for surface separations between 0.2\,$\mu$m to 0.3\,$\mu$m
using an atomic force microscope. The measured force shows
the required periodicity corresponding to the corrugations.
It also exhibits the necessary inverse fourth power distance dependence.
The obtained results are shown to be in good agreement with
a complete theory taking into account the imperfectness of
the boundary metal. This demonstration opens new opportunities
for the use of the Casimir effect for lateral translation in
microelectromechanical systems.}}


The archetypical Casimir force \cite{1} leads to an attraction between
two neutral metal plates placed in vacuum.
The force results from the alteration by the metal boundaries of
the zero point electromagnetic energy which is present in empty space.
Uniquely the Casimir force is independent of the electric
charge and other interaction constants. As a result,
it is strongly dependent on geometry and topology of
the boundary and can be both attractive as well as repulsive
(for a review of different
aspects of the Casimir effect see monographs \cite{2,3}).

The normal Casimir force which leads to an attraction perpendicular
to the two surfaces has been demonstrated first between  flat
plates \cite{4c} and between a flat plate and a lens \cite{4d,5}.
Recently extensive experimental research on the normal Casimir force
between a large sphere and a flat plate
has been performed with increased precision
\cite{6,7,8}.
Also the non-trivial boundary dependence of the Casimir force acting between
a large sphere and plate with periodic uniaxial sinusoidal corrugations
was demonstrated in Ref.~\cite{9}.
This has led to extensive theoretical study of the
corrections to the Casimir effect due to various factors as finite
conductivity
of the boundary metal, surface roughness, and nonzero temperature. Also the
combined effect of these factors was investigated (new experimental and
theoretical developments are presented in review \cite{11}). Even more
importantly, the Casimir effect is finding new applications in fundamental
science and engineering.
Thus,
with the advent of modern unified theories involving compact dimensions,
precision measurements of the Casimir force have been used to set limits
on the presence of hypothetical forces \cite{12,13,14}.
With regard to technological applications, both static and dynamic
microelectromechanical machines have been created actuated by the normal
Casimir force \cite{15,16}. Also, adhesion and sticking of moving parts in
microelectromechanical systems due to the Casimir effect were
investigated \cite{17}.

In this letter we report the first demonstration of the
lateral Casimir force.  It acts between
two aligned corrugated surfaces and leads to a mechanical force acting
tangential to the surfaces. Similar to the normal Casimir force, the
lateral Casimir force also originates from
the modifications of electromagnetic zero point oscillations.
The possibility of a
lateral Casimir force for anisotropic boundaries was investigated 
theoretically
and the harmonic dependence on a correspondent angle was predicted
(see \cite{3,18,19} for the plates made of anisotropic material and
\cite{20,21,22} for the corrugated plates made of ideal metal).
Note that other motional frictional forces between two flat parallel
surfaces have been suggested but they would be several orders of
magnitude smaller \cite{22a,22b}.
Here we
compare the experimental data with a complete theory taking into account the
finite conductivity corrections of the boundary metal. The
predicted sinusoidal dependence of the lateral force on the phase shift
and inverse fourth power dependence on surface separation is
confirmed and good agreement with a complete theory is demonstrated.
This demonstration opens new opportunities for the actuation of lateral 
motion in
microelectromechanical systems based entirely on the vacuum effects of 
quantum
electrodynamics.

The regularized zero point energy per unit area for two parallel
plates of infinite conductivity a distance $z$ apart is given by
$U(z)=-\pi^2\hbar c/(720{z}^3)$ \cite{1,2,3,11}.
This results in the normal Casimir force per unit area
$F=-\partial U/\partial z = -\pi^2\hbar c/(240{z}^4)$.
As metals have only a finite conductivity, corresponding corrections
can be incorporated in terms of the plasma wavelength $\lambda_p$
leading to  \cite{22c,22d,23}
\beq
E_{pp}(z)=-\frac{\pi^2 \hbar c}{720{z}^3}
\left[1+
\sum\limits_{n=1}^{4}
c_n\left(\frac{\lambda_p}{2\pi z}\right)^n
\right],
\label{1}
\eeq
\ni
where the coefficients are $c_1=-4$, $c_2=72/5$,
\[
c_3=-\frac{320}{7}\left(1-\frac{\pi^2}{210}\right),\quad
c_4=\frac{400}{3}\left(1-\frac{163\pi^2}{7350}\right).
\]
\ni
Eq.(\ref{1}) is applicable at the separations $z\geq\lambda_p$
with an accuracy of about (1--2)\%.

Now let us consider plates covered by
the longitudinal corrugations described by the functions
\beq
z_1=A_1\sin (2\pi x/\Lambda),\quad 
z_2=z+A_2\sin(2\pi x/\Lambda+\varphi),
\label{3}
\eeq
\ni
where $z$ is the mean separation distance between the two surfaces, 
$\Lambda$ is the 
corrugation period, $A_1$ and $A_2$ are the corrugation amplitudes. 

The normal separation distance between two points of the corrugated
surfaces can be represented as
\beq
z_2-z_1=z+b\cos(2\pi x/\Lambda-\alpha),\
\label{4}
\eeq
\ni
where
\bes
&&
b=b(\varphi)=\left(A_1^2+A_2^2-2A_1A_2\cos\varphi\right)^{1/2},
\nn\\
&&\tan\alpha =(A_2\cos\varphi -A_1)/(A_2\sin\varphi).
\nn
\ees

For corrugated plates with large corrugation period
$(\Lambda >z)$ the Casimir energy density can be found by the additive
summation of the results obtained for plane plates
(see Eq.~(\ref{1})) using the fact that all the separation distances
$z_2-z_1$ given by Eq.~(\ref{4}) are equally probable
\cite{11,22,24}
\beq
E_{pp}^{cor}(z,\varphi)=
\frac{1}{\Lambda}
\int_{0}^{\Lambda}E_{pp}(z_2-z_1)\,dx.
\label{6}
\eeq

Experimentally it is hard to maintain two parallel plates uniformly
separated by distances less than a micron. So 
one of the plates is replaced by a metal sphere of radius
$R$ where $R\gg z$.
Here a sphere of radius $R=100\,\mu$m imprinted by sinusoidal corrugations
was used instead of one of the plates
(see Fig.~1). As both $z\ll R$ and
$\Lambda\ll R$,
the normal Casimir force
can be calculated by use of the Proximity
Force Theorem \cite{25} as
$F^{nor}(z,\varphi)=2\pi RE_{pp}^{cor}(z,\varphi)$.
The accuracy of this theorem for our parameters is of order 0.2\% 
(note that the recent result \cite{25a} claiming a worse accuracy for
the PFT is applicable only to the pure nonrelativistic regime with
separations $z$ no larger than several nanometers and also small $R$).
By integrating the normal force with respect to separation, the energy
of a corrugated sphere and a plate is obtained.
Then, differentiating with respect to the phase shift, one finds
the lateral Casimir force
\beq
F^{lat}(z,\varphi)=-\frac{4\pi^2R}{\Lambda^2}
\frac{\partial }{\partial\varphi}
\int_{z}^{\infty}dy
\int_{0}^{\Lambda}E_{pp}(z_2-z_1)\,dx.
\label{8}
\eeq
\ni
Here all integrals can be calculated explicitly by the use of
representation (\ref{4}) for the separation distance.
The result is
\beq
F^{lat}(z,\varphi)=\frac{\pi^4 R\hbar c}{120z^4}
\frac{A_1A_2\sin\varphi}{\Lambda(1-\beta^2)^{5/2}}
\left[1+
\sum\limits_{n=1}^{4}c_{n,x}\!\left(
\frac{\lambda_p}{2\pi z}\right)^{\! n}
\right],
\label{9}
\eeq
\ni
where $\beta\equiv b(\varphi)/z$ and the coefficients
are given by
\bes
&&
c_{1,x}=\frac{4+\beta^2}{3(1-\beta^2)}c_1,
\quad c_{2,x}=\frac{5(4+3\beta^2)}{12(1-\beta^2)^2}c_2,
\nn \\
&&
c_{3,x}=\frac{8+12\beta^2+\beta^4}{4(1-\beta^2)^3}c_3,\quad
c_{4,x}=\frac{7(8+20\beta^2+5\beta^4)}{24(1-\beta^2)^4}c_4.
\nn
\ees
\ni
There are also corrections due to the nonzero temperature
which can be neglected for the separations reported here \cite{11}.
It can be shown that the impact of surface roughness on
$F^{lat}$ is insignificant.

The experiments are performed using a standard Atomic Force Microscope (AFM)
at a pressure below 50\,mTorr and at room temperature. 
A schematic diagram of
the experiment is shown in Fig.~1. To implement this experiment a
diffraction grating with an uniaxial sinusoidal corrugations of period
$\Lambda=1.2\,\mu$m
and an amplitude of 90\,nm was used as the template. In order to
obtain perfect orientation and phase between the two corrugated surfaces a 
special procedure was developed, where  
the in situ pressure imprinting of the corrugations 
from the plate onto the gold coated sphere was done. 
This imprinting procedure required special 
adaptation of the
cantilever.  First a polystyrene sphere was attached to the 
cantilever tip with conductive silver epoxy.  Next 
a $<10\,\mu$m thick, $100-200\,\mu$m wide and 0.5\,mm long piece
of freshly
cleaved mica is attached to the bottom of the sphere 
with silver
epoxy. Then a second polystyrene sphere of 200$\pm4\,\mu$m
diameter was mounted on the tip of
mica with the same silver epoxy.  This second sphere is 
used in the measurement.  The cantilever (with mica plate and 
spheres), corrugated 
plate
and a smooth flat plate (polished sapphire) 
were then coated with about 400 nm of gold in a thermal
evaporator.  A small section of the corrugated plate is also coated with
100\,nm of aluminum (as $Al$ is much harder than gold the imprinting of the
corrugations on the sphere is done on this surface).  The sphere and the
plates are mounted as shown in Fig.~1.  As the AFM cantilever is only
320\,$\mu$m long, the addition of the mica plate
and the second sphere  is necessary
to (i) allow access to the central regions of the corrugated plate (ii)
reduce the effect of scattered laser light from the edges of the
corrugated plate and (iii) increase the sensitivity of the 
cantilever.   Now the sphere is moved over to the region of the
corrugated surface coated with $Al$ and the corrugations
are imprinted on the gold
coating of the surface by pressure using the piezos shown.  
After this imprinting, the sphere is
translated  over to the gold coated area of the plate. Extreme 
care to preserve the parallel orientation of the two corrugations 
is necessary during this translation, as even a misallignment of 
0.1 rad can lead to crossovers of the axis of the two corrugations and 
destruction of any lateral Casimir force. We used 
optical beams to monitor the orientation of the 
cantilever holder with an accuracy of 2$\times 10^{-3}$ radians 
during the translation process. The amplitude of
the corrugation on the plate and the imprinted one on the sphere were
measured using the AFM, to be
$59\pm 7\,$nm and $8\pm1\,$nm respectively. This measurement 
was made after completion of the experiment. 

      The experimental arrangement in 
Fig.~1 is so chosen to increase the 
sensitivity for detection of lateral 
forces and suppress the effect of the 
normal Casimir force on the cantilever. Thus a 
lateral force tangential 
to the corrugated sphere surface 
would
result in the usual bending of the cantilever in response 
to the force. Whereas a force acting normal to the 
sphere and corrugated plate (from the normal Casimir force) 
would lead to the torsional deflection (rotation) 
of the cantilever. 
The torsional spring constant of this cantilever $k_{tor}$ 
is much greater than the bending spring constant $k_{ben}$, 
making it much more sensitive to detecting the 
lateral Casimir force, while simultaneously 
suppressing the effect of the normal Casimir force. 
The calibration of the cantilever 
( $k_{tor}$ and $k_{ben}$)
and the measurement of the residual potentials between the sphere and plate 
is done by electrostatic means in a manner similar to that
in Refs.~\cite{6,7,8,9}. Here the sphere is kept grounded and various 
voltages are applied to the plates.  The normal electrostatic force
between the corrugated sphere and plate is given by:
\beq
F^{el}(z,\varphi)=-\pi R\varepsilon_0
\frac{(V_1-V_0)^2}{z}
\frac{1}{\sqrt{1-\beta^2}},
\label{11}
\eeq
\ni
where as before $z$ is the total
distance between the zero corrugation levels on the 
corrugated surfaces. $V_1$ are the voltages 
applied on the corrugated plate and $V_0$ 
is the residual potential on the grounded sphere. 
If $V_1$ is applied 
to the corrugated plate, the electrostatic force on the sphere 
leads to the torsional rotation of the cantilever. From the 
electrostatic force at different 
$V_1$ we measure the torsional spring constant 
$k_{tor}=0.138\pm 0.005\,$N/m and ${V_0}=-0.135\,$V.  Similarly 
if the sphere is brought close to the bottom smooth plate 
and voltages $V_1$ are
applied to the bottom plate (here in Eq.~(\ref{11}) $ A_{1}=A_{2}=\beta=0$ 
due to the smooth surfaces), 
the electrostatic force 
leads to the normal bending of the cantilever. From the force 
at different voltages 
$V_1$ we again measure the normal spring constant 
$k_{ben} =0.0052\pm 0.0001\,$N/m. 
Note that $k_{tor}\gg k_{ben}$ is required 
for isolation and detection of the lateral Casimir force. 
The piezo extension
with applied voltage was calibrated by optical interferometry \cite{26}.

Next the lateral Casimir force is measured. The residual potential 
difference between the corrugated sphere and plate is compensated 
by application of voltage $V_0$ to the corrugated plate. The corrugated 
plate is mounted on two piezos that allow independent movement of the 
plate in the vertical and the horizontal directions 
with help of a $x$-piezo and 
$z$-piezo respectively. Movement in the 
$x$ direction with the $x$-piezo is necessary to achieve lateral phase 
shift $\varphi$ 
between the 
corrugations on the sphere and the plate.  Independent movement in the  
$z$ direction is necessary for control of the surface separation between 
the corrugated sphere and plate.  Also, the corrugated plate 
could be mounted with 
a small but non-zero tilt away from the vertical. 
This would lead to changes 
in surface separation during lateral translation of the 
plate in the $x$ direction.  To prevent this 
a small correction voltage is applied to the $z$-piezo 
synchronous with the lateral translation in $x$ direction, 
to keep the 
surface separation distance constant.

At the start of lateral force measurement,  
the sphere is brought close to the corrugated plate and the 
separation distance is kept fixed. To measure the 
lateral force $ F^{lat}$ as a function 
of the phase $\varphi$,  
the corrugated 
plate is moved in the $x$-direction in 0.46\,nm steps using the 
$x$-piezo and the lateral force is measured at each step. 
This is repeated 60 times and the average 
lateral force at each step is recorded. The average 
lateral force measured is shown as the 
solid squares in Fig.~2.  The
sinusoidal oscillations expected from Eq.~(\ref{9}) as a 
function of the phase difference between the two 
corrugations are clearly observed. Also the periodicity 
of the lateral force oscillation is in agreement with 
corrugation period of the plate.   
A sine curve fit to the observed data is shown as the 
solid line and corresponds to an amplitude of 
3.2$\times 10^{-13}$N. 
{}From Eq.~(\ref{9}) this corresponds to a separation distance of 
$z=221\pm$2\,nm between the two corrugated surfaces.
The mean quadratic error of the average lateral force amplitude
is $\sigma_{\bar{A}}=0.15\times 10^{-13}\,$N. Systematic error
is around 5\% leading to the absolute error
$\Delta_{A}^{(s)}=0.16\times 10^{-13}\,$N. Using the value of Student
coefficient $t_{0.95,\,60}=2$ one obtains for the confidence
interval $\Delta_{A}=0.46\times 10^{-13}\,$N with a 95\% confidence
probability. The resulting precision of the amplitude measurement is 
around 15\%.
 
Next the separation between the sphere and 
corrugated plate is increased by 12\,nm with the $z$-piezo 
and the measurement is repeated. 
The average measured amplitude of 
lateral force of $2.6\times 10^{-13}\,$N corresponds from Eq.~(\ref{9})
to $z=233\pm$2\,nm consistent with the 12\,nm increa\-se in 
the separation distance. Thus the measured lateral force 
is in agreement with the complete theory taking into account 
the conductivity corrections. 
The separation distance is  
increased in 12\,nm steps and the lateral force is 
measured two more times. The amplitude of the measured 
forces was found to be consistent with the corresponding separation 
distances.  In Fig.~3 a log-log plot of the 
amplitude of measured lateral force 
as a function of the separation distance is shown as 
solid squares.  Here the separation distance of 221\,nm 
determined above is used for the closest point. 
The remainder of the points are fixed by the 12\,nm step 
increase 
in the separation distance.  
A linear fit to the data yields a 
slope of 4.1$\pm$0.2 consistent with the inverse
fourth power $z$ dependence of the lateral force 
expected from Eq.~(\ref{9})
(corrections to this dependence are rather small as 
the maximal value of $\beta=0.3$). Note that a lateral force resulting 
from electrostatic effects would lead to an inverse 
second power $z$ dependence.  

In conclusion, we have made the first measurement of  
the lateral Casimir 
force.  It was measured between 
a corrugated plate and corrugated sphere using an AFM. Exact allignment  
of the corrugations on the sphere and plate was achieved 
by a special imprinting technique.  
The measured lateral Casimir 
force is sinusoidal with a periodicity corresponding 
to the corrugation period. It 
was shown to have an inverse fourth power distance 
dependence 
which is in agreement with 
the theory taking into account the conductivity corrections. 
This demonstration of the lateral Casimir force
should open new opportunities for the application
of the Casimir effect in microelectromechanical systems.

This work is supported by a National Science Foundation Nanoscale 
Exploratory Research grant and
the National Institute 
for Standards and Technology through a Precision 
Measurement Grant. G.L.K. and V.M.M. were also supported 
by CNPq.   

\newpage
\noindent
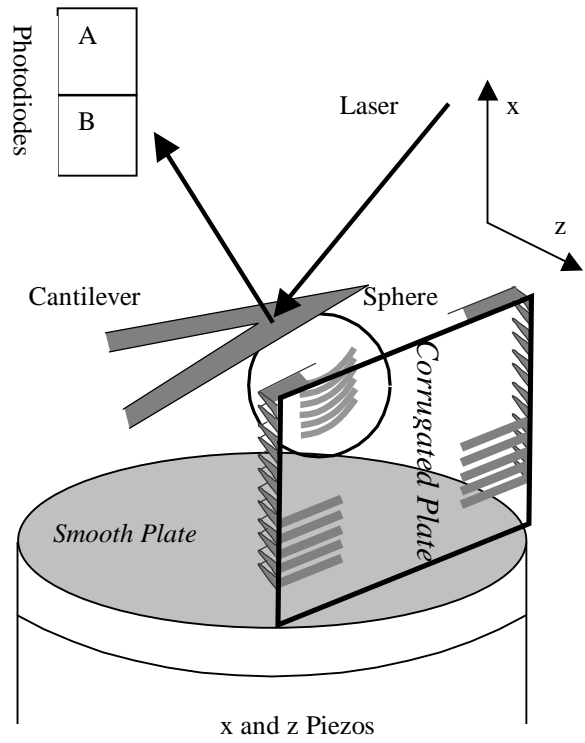
{\bf Figure 1:}  Schematic of experimental 
setup. For clarity, the size of the 
corrugations have been exaggerated 
and the additional mica plate and the first  
sphere attached to the cantilever 
are not shown. The $x$-piezo and $z$-piezo are 
independent. 
\hfill\\[5mm]
{\bf Figure 2:}  The average measured lateral 
Casimir force as a function of the lateral 
displacement of the corrugated plate is 
shown as solid squares. The solid line 
is the best fit sine curve to the data 
leading to a lateral force amplitude 
of $3.2\times 10^{-13}\,$N.
\hfill\\[5mm]
{\bf Figure 3:}  log-log plot of the measured 
lateral Casimir force amplitude as a 
function of the surface separation 
distance  is shown as solid squares. 
The slope of the straight line fit is 4.1$\pm$2.  

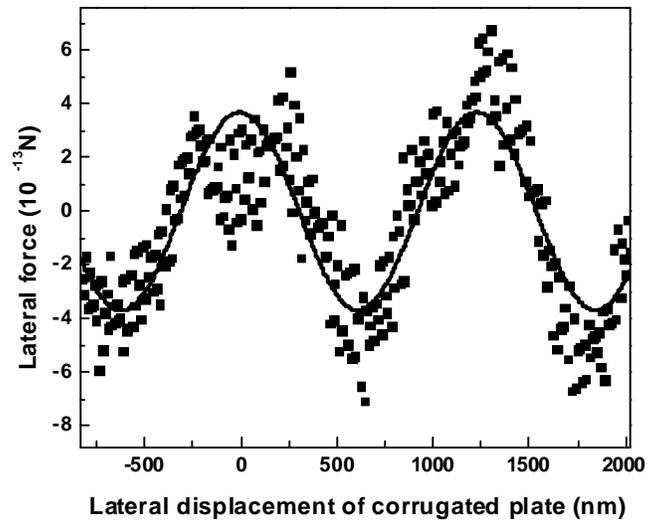
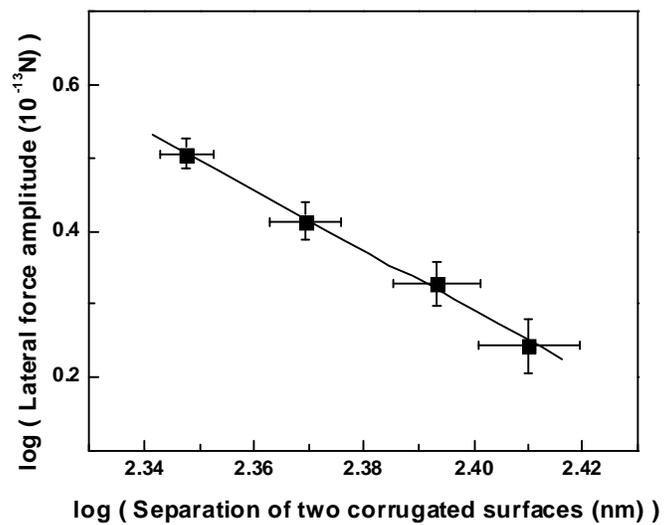
\end{document}